\def\year{2023}\relax
\documentclass[letterpaper]{article} 
\usepackage{aaai23}  
\usepackage{times}  
\usepackage{helvet} 
\usepackage{courier}  
\usepackage[hyphens]{url}  
\usepackage{graphicx} 
\usepackage{natbib}  
\usepackage{caption} 

\usepackage{subcaption}
\usepackage{tabularx}
\usepackage{booktabs}
\usepackage{comment}
\usepackage{multirow}
\usepackage{amsmath}
\usepackage{soul}
\usepackage{xspace}
\usepackage{xcolor,colortbl}

\usepackage{xcolor}

\definecolor{Gray}{gray}{0.85}
\definecolor{LightCyan}{rgb}{0.88,1,1}

\newcolumntype{a}{>{\columncolor{Gray}}r}
\nocopyright
\usepackage{xcolor}

\title{Machine-Made Media: Monitoring the Mobilization of Machine-Generated Articles on Misinformation and Mainstream News Websites}
\author {
    Hans W. A. Hanley and 
    Zakir Durumeric\\
}
\affiliations {
    Stanford University \\
    hhanley@cs.stanford.edu, zakir@cs.stanford.edu
}

\begin{document} 
\maketitle 

\begin{abstract}
As large language models (LLMs) like ChatGPT have gained traction, an increasing number of news websites have begun utilizing them to generate articles. However, not only can these language models produce factually inaccurate articles on reputable websites but disreputable news sites can utilize LLMs to mass produce misinformation. To begin to understand this phenomenon, we present one of the first large-scale studies of the prevalence of synthetic articles within online news media. To do this, we train a DeBERTa-based synthetic news detector and classify over 15.46 million articles from 3,074~misinformation and mainstream news websites. We find that between January 1, 2022, and May 1, 2023, the relative number of synthetic news articles increased by 57.3\% on mainstream websites while increasing by 474\% on misinformation sites. We find that this increase is largely driven by smaller less popular websites. Analyzing the impact of the release of ChatGPT using an interrupted-time-series, we show that while its release resulted in a marked increase in synthetic articles on small sites as well as misinformation news websites, there was not a corresponding increase on large mainstream news websites.
\end{abstract}


\section{Introduction}
Since the release of ChatGPT in November~2022, hundreds of millions of Internet users have used the large language model (LLM) to efficiently compose letters, write essays, and ask for advice~\cite{Hu2023}. However, LLMs have also been shown to produce erroneous text. In one example, CNET, a reputable website that publishes reviews and news on consumer electronics, published articles generated by OpenAI's ChatGPT that were rife with factual errors~\cite{Leffer2023}. Beyond inaccurate text, recent research has shown LLMs can be used to effectively spread misinformation~\cite{tang2023science}. Yet, despite the widespread adoption of LLMs and their potential to accelerate the spread of misinformation, there has not been any study of whether LLMs like ChatGPT \textit{have been} broadly used to produce news articles on mainstream or fringe/unreliable websites.

In this work, we present a large-scale study of the \emph{relative increase} in  \textit{machine-generated/synthetic} articles from 3,074~news websites (1,059 misinformation/unreliable websites and 2,015 mainstream/reliable news websites) between January 1, 2022, and May 1, 2023. To do this, we utilize data from 19~models, as well as adversarial data from article perturbation/re-writes and paraphrases, to train a DeBERTa-based model~\cite{he2021deberta} to detect English-language synthetic news articles. We subsequently benchmark this classifier on eight test sets of machine-generated news articles, including two from real-world companies~\cite{pu2022deepfake} and one from an independent collection of human-written real-world articles. Across these test datasets, our model, at a false positive rate (FPR) of 1\%, achieves an average precision score of 0.992. With this trained model, we classify over 15.46M~articles published between January 1, 2022, and May 1, 2023, from our set of 3,074~news websites.\footnote{We release the weights of our model and the URLs used in this study at \url{https://github.com/hanshanley/machine-made-media}.}

We find that among reliable/mainstream news websites, synthetic articles increased in prevalence by 57.3\% (0.88\% of news articles in January 2022 to 1.39\% in May 2023) while among unreliable/misinformation websites, the prevalence increased by 474\% (0.39\% of news articles in January 2022 to 2.22\% in May 2023). Examining the content of synthetic articles, we find that while mainstream/reliable news websites have largely utilized synthetic articles to report on financial and business-related news, misinformation/unreliable news websites have reported on a wide range of topics ranging from world affairs (\textit{e.g.}, the Russo-Ukrainian War) to human health (\textit{e.g.}, COVID-19). Examining the impact of ChatGPT on the prevalence of synthetic content, we further find that its release coincided with significant increases in machine-generated articles on misinformation websites and unpopular mainstream news websites.

Our work presents one of the first in-depth analyses of the growth of synthetic articles across the news ecosystem. We show that throughout 2022 and 2023, particularly after the release of ChatGPT, many misinformation websites have rapidly increased the amount of synthetic content on their websites. As misinformation websites increasingly utilize synthetic articles, we hope that our work can serve as the basis for identifying the use of LLMs and for helping enable future studies on the spread of misinformation.

\section{Background and Related Work\label{sec:related}}

Recent advances in large language models (LLMs) have resulted in impressive performance on a variety of tasks, most notably convincing text generation~\cite{brown2020language, chowdhery2022palm, ChatGPT2022, zellers2019defending}. Within the past year, models like Open AI's ChatGPT, Meta's LLaMa, and Google's Bard have largely democratized their use. However, despite their popularity, the widespread availability of LLMs can be problematic. For example, Zeller et~al.~\shortcite{zellers2019defending} showed that even the older GPT-2 LLM can create convincing articles, often with factual errors, that evoke more trust than human-written articles.

\vspace{3pt}
\noindent
\textbf{Definition: Synthetic Articles} Within this work, we consider new articles largely generated by LLMs and other automated software to be \textit{synthetic/machine-generated}~\cite{gagiano2021robustness}. For instance, the article produced by prompting the API for OpenAI's \texttt{GPT-3.5 davinici} LLM would be considered \textit{synthetic}. We note, however, as shown in prior work~\cite{mitchell2023detectgpt,uchendu2021turingbench}, heavily human-edited machine-generates news articles are difficult to detect, often being indistinguishable from human-written news articles. As such, within this work, we further define \textit{synthetic} news articles as those that are largely if not completely generated by LLMs without significant human modification. 

\vspace{3pt}
\noindent
\textbf{Real-World Use of Synthetic News Media.} While the large-scale democratization of generative models is new, the use of machine-generated or \textit{synthetic} articles by news websites is not. Since as early as 2019, Bloomberg has used the service Cyborg to automate the creation of nearly one-third of their articles~\cite{Peiser2019}. Similarly, since 2019, as reported by the New York Times, other reputable news sources including The Associated Press, The Washington Post, and The Los Angeles Times, have used machine-generation services to write articles on topics that range from minor league baseball to earthquakes~\cite{Peiser2019}. However, articles that contain machine-generated content from services such as Cyborg, BERTie, or ChatGPT, while reducing the workload of reporters, have also been shown to often contain factual errors~\cite{Alba2023,Leffer2023}. As a result, much research has focused on detecting machine-generated news articles~\cite{zellers2019defending,uchendu2020authorship,he2023mgtbench,ippolito2020automatic}.

\vspace{3pt}
\noindent
\textbf{Detecting Machine-Generated Media.} Several approaches have been developed to detect machine-generated text. BERT-defense~\cite{ippolito2020automatic} for instance uses a BERT-based~\cite{devlin2019bert} model to identify machine-generated texts. DetectGPT~\cite{mitchell2023detectgpt} approximates the probabilistic curvature of specific LLMs for zero-shot detection. Mitchell et~al.\ show that if the specific model used to generate text is known and can be readily queried to obtain the log probabilities of pieces of text, then it is possible to easily differentiate synthetic articles from human-written news articles, achieving a  0.97 AUROC for the XSum dataset. Zhong et~al.~\shortcite{zhong2020neural} propose a graph-based approach that considers the factual structure of articles to detect machine-generated text.

Our work depends on accurately identifying machine-generated articles across news websites. As shown in previous works, however, many machine learning models trained to detect synthetic texts overfit to their training domain, the token distribution of the model used to generate the synthetic texts, and the topics that they were trained on~\cite{mitchell2023detectgpt,uchendu2020authorship,lin-etal-2022-truthfulqa}. For example, models trained to detect synthetic news articles, often fail to detect shorter machine-generated tweets. Despite these shortcomings, as illustrated by Pu et al.~\shortcite{pu2022deepfake}, classifiers focused on only one domain can often perform exceedingly well on datasets seen ``in-the-wild.'' Adversarially training a RoBERTa~\cite{liu2019roberta} based classifier, Pu et~al. achieve an $F_1$-classification score of 87.4–91.4 on a test dataset made up of synthetic news articles purchased from AI Forger and Article Forge. Unlike in other domains, such as tweets or comments, news articles tend to be longer, allowing for greater precision in their classification~\cite{pu2022deepfake,sadasivan2023can}.

\vspace{2pt}
\noindent
\textbf{Reliable and Unreliable News Websites.}
In this work, we analyze how both reliable/mainstream and unreliable/misinformation news websites have published machine-generated articles throughout 2022 and 2023.  Unreliable information from these sites can take the form of \textit{misinformation}, \textit{disinformation}, and \textit{propaganda}, among others~\cite{jack2017lexicon}. Within this work, we refer to websites that have been labeled by other researchers as generally spreading \textit{false} or unreliable information as misinformation/unreliable news websites (including both websites labeled as \textit{misinformation} and \textit{disinformation} within this label). As in prior work, we consider reliable/mainstream news websites as ``outlets that generally adhere to journalistic norms including attributing authors and correcting errors; altogether publishing mostly true information''~\cite{hounsel2020identifying}.


\section{Detecting Machine-Generated Articles\label{sec:detect} }

As described in Section~\ref{sec:related}, several approaches have been developed for identifying synthetic articles, with some of the most successful being transformer-based methodologies~\cite{pu2022deepfake,gehrmann2019gltr}. However, given that past models were trained to (1) only detect text from \textit{particular} models~\cite{zellers2019defending}, (2) are deeply vulnerable to adversarial attacks~\cite{pu2022deepfake}, (3) or have unreleased weights~\cite{zhong2020neural}, we design and benchmark our \emph{own} transformer-based machine-learning classifiers to identify synthetic articles in the wild. 

In addition to training three transformer architectures (BERT, RoBERTa, DeBERTa) on a baseline training dataset (detailed below), we further train these models on datasets generated from two common adversarial attacks~\cite{krishna2023paraphrasing,mitchell2023detectgpt}. To benchmark and understand the generalization of our approach, we test our new models against datasets of articles generated by two companies, AI Writer and AI Forger provided to us by Pu et al.~\shortcite{pu2022deepfake}, the Turing Benchmark~\cite{uchendu2021turingbench}, four distinct GPT-3.5 generated datasets~\cite{OpenAI2022}, and finally a dataset of human-written articles from 2015~\cite{Signal1M2016}. We now describe our training and test datasets, the architectures of our models, and finally our models' performances on our benchmarks.

\vspace{3pt}
\noindent
\textbf{Baseline Training Datasets.} To train a classifier to detect machine-generated/synthetic news articles found in the wild, we require a diverse dataset of articles from a wide array of generative models. Thus, for our baseline training dataset, we take training data of machine-generated/synthetic articles from three primary sources: the Turing Benchmark, Grover, and articles generated from GPT-3.5.

\textit{Machine-Generated Training Articles:} For much of our training data, we utilize the Turing Benchmark~\cite{uchendu2021turingbench}, which contains news articles generated by 10~different generative text architectures including GPT-1~\cite{radfordimproving}, GPT-2~\cite{radford2019language}, GPT-3~\cite{brown2020language}, CTRL~\cite{keskar2019ctrl}, XLM~\cite{lample2019cross}, Grover~\cite{zellers2019defending}, XLNet~\cite{yang2019xlnet}, Transformer-XL~\cite{dai2019transformer}, and FAIR/WMT~\cite{ng2019facebook,chen2020facebook}. We note that given the different settings and trained weights provided by the authors of these respective works, the Turing Benchmark altogether includes articles generated from 19 different models. We randomly subselect 1000~articles from within the Turing benchmark generated by each of these different models as training data.

In addition to the Turing Benchmark training dataset, we use the training dataset of Zellers et~al.~\shortcite{zellers2019defending}, which contains realistic, often long-form articles, that mimic the fashion of popular news websites such as cnn.com, nytimes.com, and the washingtonpost.com. Unlike the Grover-generated articles from the Turing Benchmark dataset, which are generated using a prompt of just the title of potential articles, these Grover articles are generated in an unconditional setting \textit{and} from prompting the Grover model with metadata (\textit{i.e.}, title, author, date, website). As found by Zellers et~al., many of the articles produced by their models were convincing to human readers, and we thus include 11,930~machine-generated articles from the base model of Grover (across different Grover decoding settings [\textit{e.g.}, p=1.00, p= 0.96, p=0.92 (nucleus/top-p), k=40 (top-k), \textit{etc...} settings ]) in our training dataset.

Finally, given the popularity of the GPT-3.5 model~\cite{Hu2023}, with it being the basis of the released version of ChatGPT, and GPT-3.5 being one of the most powerful released models, we add 3,516~articles generated from the \texttt{GPT-3.5 davinici} model. To create these articles, we prompt the public API of \texttt{GPT-3.5 davinici} with the first 10 words of 3,516 real news articles from 2018 (see Section~\ref{sec:dataset}; while scraping our news dataset, we acquired several million articles from 2018). For \texttt{GPT-3.5 davinici} model, we use a nucleus decoding setting of p=1.00, p=0.96, and p=0.92 (some of the most common~\cite{mitchell2023detectgpt,zellers2019defending}).

We finally note that, as found in prior work~\cite{pu2022deepfake,uchendu2021turingbench,zellers2019defending}, machine-generated news articles are often shorter in length than human-written articles. While training, to ensure that our models do not simply distinguish between longer human-written articles and those generated by generative transformers by their different lengths, we ensure that our machine-generated and human-written articles are of similar lengths (median training synthetic article length of 210 words and median training human article length of 224 words). Furthermore, as found by past work, predictions for texts, particularly short texts, tend to be unreliable~\cite{kirchner2023classifier,zellers2019defending,pu2022deepfake}; conversely, as shown by Sadasivan et al.~\shortcite{sadasivan2023can}, as the lengths of texts increase the variance between human and machine-generated texts increases. As such, for our training and our generated test data (GPT 3.5 dataset), we exclude texts shorter than 1,000~characters (140 words)~\cite{OpenAI2022}. We note as a result, we do not use every trained model's articles from the Turing Benchmark; given that WMT-20/FAIR articles within this dataset are all shorter than 1000~characters, we do not include them within our training dataset. Altogether our training dataset thus includes data from 19 different models (18 from Turing Benchmark and \texttt{GPT-3.5 davinici}). 

\textit{Human-Written Training Articles:} For our set of human-generated articles, as in Zellers et~al.~\shortcite{zellers2019defending}, we utilize news articles published in 2018. Specifically, we use 28,446~articles from 2018 from our set of news websites that we later measure (see Section~\ref{sec:dataset}; while scraping our news dataset, we acquired several million articles from 2018), 2,500~articles from the human split of the Grover dataset, and 2,500 articles from the human-train-split within the Turing Benchmark dataset. 

We present an overview of our complete \texttt{Baseline} dataset in Table~\ref{tab:training-dataset}.


 





\begin{table}

\centering

\footnotesize

\begin{tabular}{l|rr}


\textbf{} &\textbf{Human } & \textbf{Machine} \\

\textbf{Training Dataset} & \textbf{Written}& \textbf{Generated}\\\midrule

Baseline & 33,446 & 33,446 \\

Pert. & 33,446 & 44,003 \\

Para. & 33,446 & 41,498 \\

Perturb + Para. & 33,446 & 52,055 \\

\bottomrule

\end{tabular}

\caption{\label{tab:training-dataset} The number of machine-generated and human-written articles within the \texttt{Baseline}, \texttt{Pert}, \texttt{Para}, and \texttt{Pert.+Para.} training datasets.}

\vspace{-10pt}

\end{table}

\begin{table}

\centering

\footnotesize

\begin{tabular}{l|rr}

\textbf{} &\textbf{Human } & \textbf{Machine} \\

\textbf{Test Dataset} & \textbf{Written}& \textbf{Generated}\\\midrule

Turing Benchmark & 975 & 18,076 \\

GPT-3.5 & 1,000 & 243 \\

GPT-3.5 w/ Pert. & 1,000 & 241\\

GPT-3.5 w/ Para. & 1,000 & 118\\

Article Forger & 1,000 & 1,000\\

AI Writer & 1,000 & 1,000 \\

\bottomrule

\end{tabular}

\caption{\label{tab:test-dataset} The number of machine-generated and human-written articles within our test datasets. }

\vspace{-10pt}

\end{table}

\begin{table*}

\centering


\fontsize{6.4pt}{6}

\selectfont

\setlength\tabcolsep{4pt}

\begin{tabular}{l|ccc|ccc|ccc|ccc|ccc|ccc|c}

& \multicolumn{3}{c|}{\textbf{Turing Benchmark}} & \multicolumn{3}{c|}{\textbf{GPT-3.5 }} & \multicolumn{3}{c|}{\textbf{GPT-3.5 w/ Pert}} &\multicolumn{3}{c}{\textbf{GPT-3.5 w/ Para } } &\multicolumn{3}{c}{\textbf{Article Forger}} & \multicolumn{3}{c}{\textbf{AI Writer}} &\textbf{ Avg.} \\

& \text{$F_1$} & Prec. & Recall & \text{$F_1$} & Prec. & Recall & \text{$F_1$} & Prec. & Recall & \text{$F_1$} & Prec. & Recall & \text{$F_1$} & Prec. & Recall & \text{$F_1$} & Prec. & Recall &\textbf{ \text{$F_1$}}\\ \midrule

OpenAI Roberta & 0.717& 0.997 & 0.560 & 0.092 & 0.684 & 0.049 & 0.022 &0.375& 0.011& 0.309 & 0.950 & 0.185 &0.750 & 1.000 & 0.600 & 0.881 & 1.000 & 0.787 & 0.462\\\midrule


BERT & 0.988 & 0.998 & 0.978 & 0.941 & 0.911 & 0.973 & 0.901 & 0.905 & 0.898 & 0.931 & 0.889 & 0.976 & 0.855 & 0.809 & 0.905 & 0.779 & 0.929 & 0.670 & 0.899\\\midrule

BERT+ Pert. &0.995 &0.992&0.998 & 0.898 & 0.817 & 0.996 & 0.896 & 0.816 & 0.992 & 0.872 & 0.777 &0.995& 0.808 & 0.681 & 0.994 & 0.892 & 0.771 & 0.995 & 0.894\\\midrule

BERT + Para. & 0.995 & 0.992& 0.998 & 0.937 & 0.896 & 0.981 & 0.915 & 0.892 & 0.939 & 0.930 & 0.822 & 0.995 & 0.839 &0.763 & 0.931 &  0.856& 0.888 & 0.826 & 0.912 \\ \midrule 

BERT+Pert.+Para. &0.995 & 0.994 &0.997 & 0.939 & 0.897 & 0.985 & 0.937 & 0.896 & 0.981 & 0.925 & 0.871 & 0.985 & 0.854 & 0.903 & 0.913 & 0.809 & 0.909 & 0.729 & 0.910\\\midrule 

RoBERTa & \textbf{0.998} & 0.997 & 0.998 & 0.956 & 0.929 & 0.985 & 0.952 & 0.928 & 0.977 & 0.949 & 0.911 & 0.990 & 0.856 & 0.857 & 0.872 & 0.951 & 0.933 & 0.971 & 0.943\\\midrule 

RoBERTa + Pert. & 0.993 &\textbf{1.000} &0.986 & \textbf{0.979} & 0.981 & 0.977 &\textbf{0.975} & \textbf{0.981}&0.977 &0.968& \textbf{0.975} & 0.961 & 0.748 & \textbf{0.977} & 0.606&  0.820 & \textbf{0.997} & 0.696 & 0.914\\\midrule 

RoBERTa + Para &\textbf{0.998} &0.998 &0.998 & 0.940 & 0.902 & 0.981 & 0.932 & 0.901 & 0.966& 0.934& 0.880 & 0.995 & 0.903 & 0.849 & 0.965 & 0.958 & 0.927 & 0.991 &0.944 \\\midrule

RoBERTa+Pert.+Para. & 0.995 & 0.991 & 0.999 & 0.956 &0.923 & 0.992 & 0.960 & 0.923& 1.000 & 0.947 & 0.903 & 0.995 & 0.912 & 0.859 & 0.972 & 0.951 & 0.913 & 0.991 & 0.954 \\\midrule  

DeBERTa &0.995 & 0.997& 0.993 & 0.961 & 0.935 & 0.989 & 0.952 & 0.934 & 0.970 & 0.958 & 0.920 & \textbf{1.000} & 0.959 & 0.951 & 0.968 & 0.986 & 0.982 & 0.989  &0.969\\\midrule  

DeBERTa + Pert. & 0.996 & 0.995 & 0.996 & 0.943 &0.895& 0.996 & 0.945 & 0.895 & \textbf{1.000} & 0.930 & 0.869 &\textbf{1.000} & 0.956 & 0.927 & 0.987 & 0.985 & 0.975 & {0.996} &0.959\\\midrule  

DeBERTa + Para. & 0.996 &0.993 & 0.999 &0.941& 0.892&0.996 & 0.943 & 0.892 & \textbf{1.000}& 0.928 & 0.866 & 1.000 & 0.965 & 0.940 & \textbf{0.991} & 0.983 & 0.967 & 1.000 &0.959\\\midrule 

DeBERTa+Pert.+Para. &0.995&0.994&0.996 & {0.970} & 0.949 & 0.992 & 0.972 &0.949&0.996&0.967 & 0.936 & \textbf{1.000} & \textbf{0.968} & 0.948& 0.989 &\textbf{0.990} & 0.979 & \textbf{1.000} & \textbf{0.977}\\\midrule  


\end{tabular}

\caption{\label{tab:benchmark} Binary \text{$F_1$}-Score/Precision/Recall of our models on various benchmarks (\textit{machine-generated/synthetic} being positive). We bold the best score in each column. As seen, our set of DeBERTa models performs the best across many of the test datasets, with \texttt{DeBERTa+Pert+Para} having the highest average F-1 score across all six datasets. }
\vspace{-10pt}
\end{table*}
\vspace{3pt}
\noindent
\textbf{Baseline Test Datasets.}
For our baseline test datasets (Table~\ref{tab:test-dataset}), we utilize the validation split from the Turing Benchmark (the labels from the test split were unavailable to us), and another test dataset consisting of 243~additional GPT-3.5 articles that we created by again prompting \texttt{GPT-3.5 davinici}, and 1000~human-written articles from 2018 (see Section~\ref{sec:dataset}; as with our training data, while scraping our news dataset, we acquired several million articles from before 2019). Further, to ensure our models generalize and handle articles seen in the wild, we utilize the \textit{In-the-Wild} dataset provided to us by Pu et~al.~\shortcite{pu2022deepfake}. This dataset consists of news articles created using generative large language models from two independent companies, Article Forger and AI Writer. By testing against these outside datasets, we validate our approach against articles generated by (1)~models not within our dataset and (2)~by generative news article services available to the public. We provide details in Table~\ref{tab:test-dataset}.

\vspace{3pt}
\noindent
\textbf{Training and Test Dataset using Perturbations and Paraphrases.} Transformer-based classifiers are often particularly susceptible to adversarial attacks, particularly attacks that rewrite sections of the generated article~\cite{mitchell2023detectgpt,pu2022deepfake} and paraphrase attacks (\textit{i.e.}, where a generic model is used to paraphrase the output of a different generative model~\cite{krishna2023paraphrasing}). To guard against these weaknesses, we take two approaches (1) perturbing our set of synthetic articles by rewriting at least 25\% of their content using the generic T5-1.1-XL model\footnote{\url{https://huggingface.co/google/t5-v1_1-large}} and (2) paraphrasing each article with the T5-based Dipper model.\footnote{\url{https://huggingface.co/kalpeshk2011/dipper-paraphraser-xxl}}



\textit{Constructing Perturbed Synthetic Articles.} To perturb/rewrite sections of our machine-generated articles, as in Mitchell et~al.~\shortcite{mitchell2023detectgpt}, we randomly MASK 5-word spans of text until at least 25\% of the words in the article are masked. Then, using the text-to-text generative model T5-3B~\cite{raffel2020exploring}, we fill in these spans, perturbing our original generated articles. As shown by Mitchell et~al.~\shortcite{mitchell2023detectgpt}, large generic generative models such as T5 can apply perturbations that roughly capture meaningful variations of the original passage rather than arbitrary edits. This enables us to model divergences from the distributions of texts created by our 19 different generative models (18 from Turing Benchmark and GPT-3.5). We thus utilize T5-3B\footnote{\url{https://huggingface.co/google/t5-v1_1-large} } to perturb a portion of the machine-generated articles of our \texttt{Baseline} train dataset. In addition, we create a separate test dataset by perturbing our GPT-3.5 test dataset  (Table~\ref{tab:test-dataset}). We note that after perturbing our datasets, we filter to ensure all articles used for training contain at least 1000~characters. We annotate training and test datasets containing synthetic articles perturbed with T5-3B with the suffix \texttt{Pert}. After perturbation we still consider these articles to be synthetic.

\textit{Constructing Paraphrased Synthetic Articles.} To paraphrase each of the machine-generated articles within our dataset, we use the approach outlined by Krishna et~al.~\shortcite{krishna2023paraphrasing}. Specifically, as in their work, we utilize {Dipper}, a version of the T5 generative model fine-tuned on paragraph-level paraphrases, that outputs paraphrased versions of the inputted text. We use the default and recommended parameters\footnote{We use a lexical diversity parameter of 60. For more details on the Dipper model see Krishna et~al.~\shortcite{krishna2023paraphrasing}} as in Krishna et~al. to paraphrase a portion of the text within our original training dataset as well as our GPT-3.5 test dataset~\cite{krishna2023paraphrasing}. We note that after paraphrasing our datasets, we again filter to ensure all articles utilized for training contain at least 1000 characters  (Table~\ref{tab:test-dataset}). We annotate training and test datasets containing articles paraphrased with Dipper with the suffix \texttt{Para}. After paraphrasing we still consider these articles to be synthetic.

\vspace{3pt}
\noindent
\textbf{Detection Models.}
Having described our training test sets, we now detail our models and evaluate their performance on our 6~test datasets (Turing Benchmark, GPT-3.5, GPT-3.5 w/ \texttt{Pert}, GPT-3.5 w/ \texttt{Para}, Article Forger, AI Writer). Specifically, we fine-tune three pre-trained transformers, BERT-base~\cite{devlin2019bert}, RoBERTa-base~\cite{liu2019roberta}, and DeBERTa-v3-base~\cite{he2021deberta}\footnote{\url{https://huggingface.co/bert-base-uncased}}\footnote{\url{https://huggingface.co/roberta-base}}\footnote{\url{https://huggingface.co/microsoft/deberta-v3-base}}. For each architecture, we train 4~models to detect machine-generated news articles using our \texttt{Baseline}, \texttt{Perturb}, \texttt{Para}, and \texttt{Perturb+Para} training datasets. For each architecture, we build a classifier by training an MLP/binary classification layer on top of the outputted [CLS] token. We use a max token length of 512~\cite{ippolito2020automatic,pu2022deepfake}, a batch size of 32, and a learning rate of $1\times10^{-5}$. Each model took approximately 2~hours to train using an Nvidia RTX A6000 GPU\@. After training, as in Pu et~al.~\shortcite{pu2022deepfake}, we determine each model's binary \text{$F_1$}-scores, precision, and recall for each test dataset and rank each model using its average \text{$F_1$}-score. We classify each text based on its outputted softmax probability ($>$0.5 being classified as \textit{synthetic}). For a baseline comparison for our trained models, we further test the Roberta-based classifier released by Open AI in 2019~\cite{solaiman2019release} on each of our test datasets.
\begin{table}

\centering


\fontsize{8.4pt}{6}

\selectfont

\setlength\tabcolsep{4pt}

\begin{tabular}{l|ccc|c|}

& \multicolumn{3}{c|}{\textbf{ChatGPT Rewrite}} & {\textbf{Signal Art. }}\\
& \text{$F_1$} & Prec. & Recall & Accuracy \\ \midrule
OpenAI Roberta & 0.002 & 0.200& 0.001 & 0.997\\ \midrule
BERT+Para& 0.905 &0.978 &  0.842 &0.766 \\\midrule

RoBERTa+Pert.+Para. & 0.937 & 0.964 & 0.912 & 0.820 \\\midrule
DeBERTa+Pert.+Para. &0.892 &  0.979 &  0.820 & 0.942 \\\midrule 


\end{tabular}

\caption{\label{tab:benchmark2} We benchmark our BERT \texttt{+Para}, RoBERTa\texttt{+Pert+Para}, and DeBERTa \texttt{+Pert+Para} models, and the OpenAI RoBERTa model on a dataset of 1,000 articles from 2018 rewritten by ChatGPT (along with the original 1,000 human-written articles) and a dataset of 10,000 human-written articles from 2015 chosen randomly from the Signal article dataset.}
\vspace{-10pt}
\end{table}

Consistent with prior works~\cite{veselovsky2023artificial,mitchell2023detectgpt,gagiano2021robustness}, due to training our model on synthetic articles from a wide variety of sources, and due to our model's focus on news articles, as seen in Table~\ref{tab:benchmark}, we observe that all our trained models perform markedly better than Open AI's 2019 released detection model. We further observe, as aggregated in the \texttt{Avg. \text{$F_1$}}-score column, our set of DeBerta models performs the best in classifying machine-generated/synthetic content, all achieving an average \text{$F_1$}-score greater than 0.959. In particular, we observe that our DeBERTa model trained on a dataset that includes our set of adversarial data \texttt{Pert} + \texttt{Para}, performs the best at an average \text{$F_1$}-score of 0.977. This particular model further achieves the best respective \text{$F_1$}-scores in classifying the set of articles from Article Forger and AI-writer provided by Pu et~al., achieving \text{$F_1$}-scores of 0.968 and 0.999 on the two datasets respectively. We note that our model, in addition to performing better than Open AI's Roberta, also outperforms all models benchmarked by Pu et al.~\shortcite{pu2022deepfake} on the  AI Forger and the AI Writer test datasets, which achieved \text{$F_1$}-scores ranging from 1.6 to 94.9. This illustrates that our model generalizes to other types of machine-generated articles from models not included in our dataset. 

In addition to testing our models on these six datasets, to further ensure that our approach generalizes well, we test our models in two additional settings: (1) a setting where ChatGPT is utilized to rewrite a given human-written article, (2) a setting that includes articles not from the year of training (2018) and from websites not in our original dataset. As such, we finally test the OpenAI Roberta classifier as well as the best BERT, RoBERTa, and DeBERTa models on (1) a set of 1,000 articles from our dataset of 2018 news articles that were rewritten\footnote{We had ChatGPT rewrite each article by supplying the prompt ``Rewrite the following news article in your own words:'' followed by the article.} by ChatGPT~\cite{OpenAI2022} as well as the corresponding set original news articles, and (2) 10,000 randomly selected human-written articles from 2015 from the Signal Media news article dataset. As seen in Table~\ref{tab:benchmark2}, our DeBERTa\texttt{+Pert+Para} model achieved the highest accuracy on the Signal dataset and the second highest precision on the ChatGPT~ Rewrite dataset, with scores of 94.2\% accuracy and 97.9\% precision respectively.

\vspace{3pt}
\noindent
\textbf{Selecting a classification threshold for \textit{synthetic} articles.}
Given its performance across all eight of our datasets, we use our \texttt{DeBERTa+Pert+Para} trained model as our detection model for the rest of this work. However, as noted in prior research~\cite{krishna2023paraphrasing}, a realistic low false positive rate (FPR) would be near 1\%. Given our model only achieves an average FPR of 5.8\% on our Signal article dataset at a softmax probability threshold of 0.50, when classifying articles within this work, we raise our softmax probability classification threshold to 0.98\%, allowing us to achieve a 1\% FPR/accuracy on the Signal article dataset. At this threshold, our model achieves a 0.993/0.972 precision/recall on our original six datasets with an FPR of 0.7\%. Similarly, at this threshold, our model reaches a precision of 0.989 on our ChatGPT rewrite test set at the expense of only reaching a 0.639 recall. We thus find that by increasing our threshold to 0.98, we can achieve a realistic FPR at the expense of recall. For the rest of this work, we utilize a softmax probability threshold of 0.98. Our work thus likely represents a conservative estimate of the amount of \textit{synthetic} articles online.


\section{News Dataset and Classification Pipeline \label{sec:dataset}}

Having described the DeBERTa-based model that we use to identify machine-generated/synthetic articles, we now describe our datasets of scraped news articles. 

\vspace{3pt}
\noindent
\textbf{Website List.} Between January 1, 2022, and May 1, 2023, we gather all articles published from 3,074~news websites.\footnote{We note that while this study focuses on the release of ChatGPT as a possible focal point, our data collection for this project actually began in January 2022.} Our list of websites consists of domains labeled as ``news'' by Media Bias Fact Check\footnote{https://mediabiasfactcheck.com/} and by prior work~\cite{hanley2023golden}. Within our list of news sites, we differentiate between ``unreliable news websites'' and ``reliable news websites.'' Our list of unreliable news websites includes 1,059~domains labeled as ``conspiracy/pseudoscience'' by mediabiasfactcheck.com as well as those labeled as ``unreliable news'', misinformation, or disinformation by prior work~\cite{hanley2023golden,iffy2022,fakenews2020}. Our set of ``unreliable'' or misinformation news websites includes websites like realjewnews.com, davidduke.com, thegatewaypundit.com, and breitbart.com. We note that despite being labeled unreliable every article from each of these websites \textit{is not} necessarily misinformation. 

Our set of ``reliable''/mainstream news websites consists of the news websites that were labeled as belonging to the ``center'', ``center-left'', or ``center-right'' by Media Bias Fact Check as well as websites labeled as ``reliable'' or ``mainstream'' by other works~\cite{hanley2023golden,iffy2022,fakenews2020}. This set of ``reliable news websites'' includes websites like washingtonpost.com, reuters.com, apnews.com, cnn.com, and foxnews.com. Altogether after removing duplicates and unavailable websites, we scraped 2,015 ``reliable news'' or mainstream websites.

We note that to later understand how websites of varying popularity/size have used machine-generated articles on their websites, we striate our list of websites by their popularity using ranking data provided by the Google Chrome User Report (CrUX)~\cite{ruth2022toppling}. We note that the CrUX dataset, rather than providing individual popularity ranks for each website, instead provides rank order magnitude buckets (\textit{e.g.}, top 10K, 100K, 1M, 10M websites). As such, we analyze our set of websites in the following buckets: Rank $<$ 10K (125~websites), 10K $<$ Rank $<$ 100K (511~websites), 100K $<$ Rank $<$ 1M (1,164~websites), 1M $<$ Rank $<$ 10M (802~websites), and finally Rank $>$ 10M (472~websites).

\vspace{3pt}
\noindent
\textbf{Article Collection.} To collect the articles published by our set of news websites, we queried each website's RSS feeds (if available) and crawled the homepages of each website daily from January 1, 2022, to May 1, 2023. Upon identifying newly published articles, we subsequently scraped websites using Colly\footnote{https://github.com/gocolly/colly} and Headless Chrome, orchestrated with Python Selenium. To extract the article text and publication date from each HTML page, we parsed the scraped HTML using the Python libraries \texttt{newspaper3k} and \texttt{htmldate}. 

Given that many of our websites (\textit{e.g.}, cnn.com) have multilingual options, we use the Python \texttt{langdetect} library to filter out all non-English articles. To prepare data for classification, we remove boilerplate language using the Python\texttt{justext} library and then remove URLs, emojis, and HTML tags. Further, to ensure the reliability of our classifications, we only classify news articles that are at least 1000~characters (approximately 140 words) long. Altogether, from our selection of 3,074~websites, we gathered 15.46M~articles (12.06M from mainstream websites and 3.39M from misinformation websites) that were published between January 1, 2022, and May 1, 2023. Finally, we utilize our \texttt{DeBERTa+Pert+Para} model at a softmax classification threshold of 0.98 to classify each article as either human-written or machine-generated. Classifying all 15.46M~articles took approximately 65.8~hours using an Nvidia RTX A6000 GPU.

\noindent
\paragraph{Ethical Considerations.}
With the rise of LLMs, many companies have widely scraped and gathered data from websites to fuel their models~\cite{Schappert2023}. As a result, websites ranging from Twitter to Reddit have begun to set up restrictions to ensure the privacy of their users and to protect their content from being used in other private companies' generative models. While we do not train a generative model that could artificially produce convincing and seemingly unique reproductions of the texts that we utilize, we note the concern that our work raises. Our work, however, only \textit{studies} the texts of our set of 15.46 million articles and classifies them as machine-generated or human-written. We do not seek to generate summaries or artificial rewrites of this content. In terms of web crawling for this data, as noted elsewhere~\cite{singrodia2019review,hanley2023golden,smith2013dirt}, website crawling and scraping remain pivotal for understanding and documenting what occurs on the Internet. Without scraping, understanding trends and how the Internet could potentially affect real life becomes impossible. As decided in Van Burn v. United States, publically accessible information can be legally scraped as long as it is done ethically and does no harm to the site~\cite{VanBuren}. As such, we collect only publicly available data from our set of websites and follow the best practices for web crawling as in Acar et al.~\shortcite{acar2014web}.  We limit the load that each news site experiences by checking for new articles daily at a maximum rate of one
request every 10 seconds. The hosts that we scan from are identifiable through WHOIS, reverse DNS, and an HTTP landing page explaining how to reach us if they would like to be removed from the study. During our crawling period, we received no requests from websites to opt out.
\section{The Rise of Machine-Generated Media\label{sec:comparing}}

\begin{figure}

\centering

\includegraphics[width=0.95\linewidth]{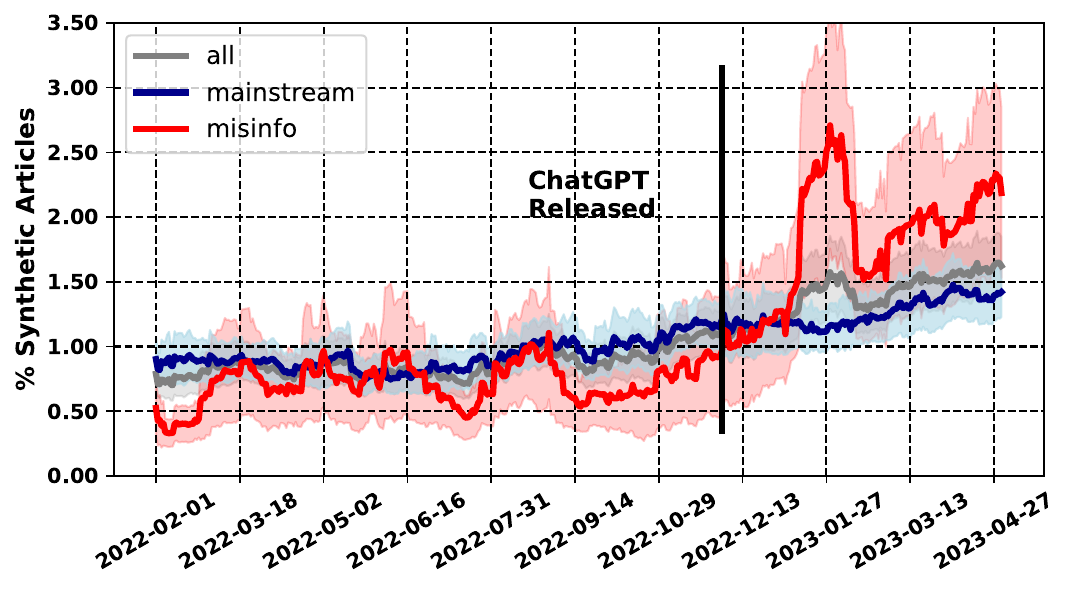}

\caption{The average percentage of synthetic articles for all, misinformation, and mainstream websites. We provide 95\% Normal confidence intervals.}

\label{fig:gpt_over_time_all}

\vspace{-15pt}

\end{figure}

Having described our detection model and datasets, in this section, we analyze the \textit{relative change} in the levels of synthetic content across our set of websites between January 1, 2022, and May 1, 2023. Specifically, we determine (1) whether there has been an increase in the use of synthetic articles, (2) if there has been an increase in their use, which sets of websites are driving this increase, (3) what synthetic articles are topically about, and (4) whether the introduction of ChatGPT has had an effect on the prevalence of synthetic articles.



\vspace{3pt}
\noindent
\textbf{Large-Scale Trends in Machine-Generated Media.} To begin, we plot the average percentage of synthetic news articles per website across our dataset between January 1, 2022, and May 1, 2023, in Figure~\ref{fig:gpt_over_time_all}. In aggregate, across all 3,074~sites, we see that 1.07\% of all articles published in January 2022 (12,984 of 1,213,983 articles) were synthetically generated. However, by May~2023, the fraction of synthetic articles nearly went up to 1.78\% (25,561 of 1,439,812~articles), a 66.0\% relative increase (nearly doubling in raw amount).

We observe that our set of reliable/mainstream websites typically had a greater percentage of synthetic articles at the beginning of 2022 compared with misinformation/unreliable news websites. While only 0.39\% of articles on average per domain from our set of misinformation websites were classified as machine-generated in January 2022, 0.88\% of articles on average from our set of mainstream/reliable websites were classified as machine-generated. This result is consistent with prior observations that many news websites have begun to use automated services to write quick, often financial-related articles (Section~\ref{sec:related}). For example, the beginning of one of the articles from Reuters (Figure~\ref{figure:machine-generated}) classified by our system as being machine-generated simply contained simple information about the direction of particular markets and funds. 


\begin{figure}

\noindent\fcolorbox{black}{lightgray}{

\begin{minipage}{.47\textwidth}

\scriptsize

\textbf{Snippet from Reuters} The S\&P 500 (.SPX) and Nasdaq (.IXIC) added to losses, while the Dow (.DJI) turned negative on Wednesday after the release of the latest FOMC meeting minutes showed that officials said the central bank may need to raise interest rates sooner than expected and reduce asset holdings quickly.

\end{minipage}}

\caption{Example first paragraph of an article classified by our system as machine-generated/synthetic.}

\label{figure:machine-generated}

\vspace{-10pt}

\end{figure}
\begin{figure}[!h]

 \centering

 \includegraphics[width=0.95\linewidth]{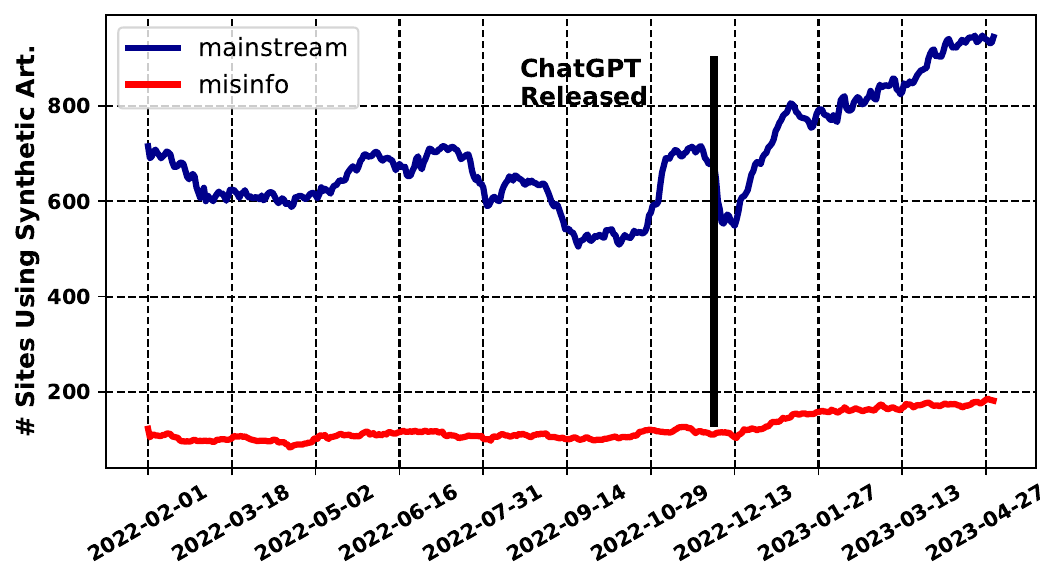}

\caption{The number of websites that published at least one synthetic article over a 30-day time span.}
\label{fig:num_using}
\vspace{-10pt}
\end{figure}
\begin{figure}[!h]

\centering

\includegraphics[width=0.95\linewidth]{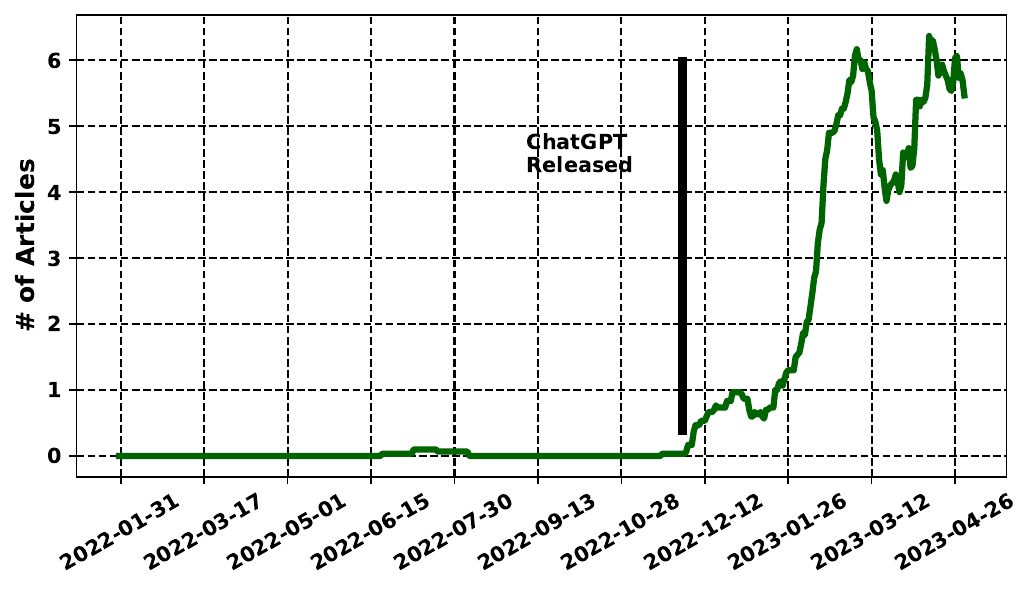}

\caption{The number of articles that contained a common ChatGPT error message over time.}

\label{fig:gpt-error}

\vspace{-10pt}

\end{figure}
\begin{figure*}
 \centering
 \includegraphics[width=1.00\linewidth]{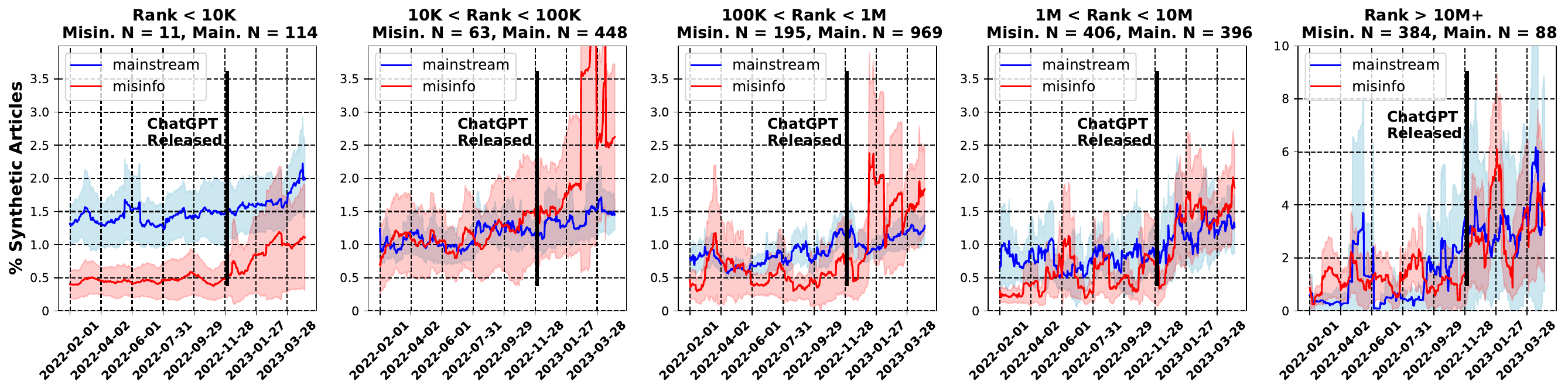}
\caption{The average percentage of machine-generated/synthetic articles for misinformation/unreliable and mainstream/reliable news websites at different striations of popularity according to Google Chrome User Report (CrUX) from October 2022. All striations of misinformation websites experienced a small uptick of machine-generated content around November 30, 2022, the release date of OpenAI's ChatGPT. \textbf{We note that the scale of synthetic content is much larger for websites with popularity rank $>$10M.}}

\label{fig:gpt_over_time}

\vspace{-10pt}

\end{figure*}

However, despite reliable/mainstream websites initially having higher levels of synthetic text, misinformation websites had marked increases in levels of machine-generated content during 2022 and 2023 (Figure~\ref{fig:gpt_over_time}). While between January 1, 2022, and May 1, 2023, reliable/mainstream news websites had a 57.3\% relative increase (0.51\% absolute percentage increase) in their levels of synthetic content, misinformation websites had a 474\% relative increase (1.85\% absolute percentage increase). Starting from a lower base, we thus see a substantial increase in the number of synthetic articles from unreliable/misinformation websites.

Furthermore, as seen in Figure~\ref{fig:num_using}, we further observe that an increasing number of news outlets published at least one synthetic article within any given 30-day time frame. Across our period of study, the number of mainstream websites that published at least one synthetic article increased from 697 (34.6\% of mainstream websites) in January 2022 to 940 (46.6\%) in April 2023. Similarly, the number of misinformation websites that published at least one synthetic article increased from 110 (10.4\% of misinformation websites) to 179 (16.9\%).

\begin{table}

\centering

\fontsize{7.5pt}{7}

\selectfont

\setlength\tabcolsep{4pt}


\begin{tabular}{l|ll|ll}

 & Misinfo Abs.  &Misinfo Rel.  & Main. Abs. & Main. Rel. \\ \toprule
Rank & \% Inc & \% Inc& \% Inc&\% Inc\\ \toprule
\textbf{All} &$1.85\%$ & $474\%$ & $0.51\%$ & $57.3\%$ \\ \midrule

Rank $<$ 10K & $0.70\%$& $175\%$ & $0.36\%$& $27.0\%$\\

10K $<$ Rank $<100$K & $1.77\%$& $221\%$ & $0.26\%$&$27.3\%$\\

100K $<$ Rank $<1$M & $1.38\%$ &$349\%$ & $0.23\%$&$30.1\%$\\

1M $<$ Rank $<10$M & $1.55\%$ &$646\%$ & $0.14\%$&$15.4\%$\\

Rank $>10$M & $3.42\%$ &$736\%$ & $2.13\%$&$423\%$\\

\bottomrule

\end{tabular}

\caption{\label{tab:increase-brackets} Estimated absolute percentage increase in machine-generated/synthetic articles between January 1, 2022 and May 1, 2023.}

\vspace{-5pt}

\end{table}
To confirm these initial findings, we further examine the increase in common idiosyncratic error messages often returned by ChatGPT. Specifically using a list of error messages including \textit{``my cutoff date in September 2021'', ``as an AI language model'', and ``I cannot complete this prompt'' } that the company News Guard~\cite{newsguard2023} has used to detect AI-generated websites, we gather every article among our 15.46 million articles that utilized such message: altogether 570 articles from 280 domains. Amongst these websites, the top domains of these articles included forbes.com (32~articles), dailymail.co.uk (29), fairobserver.com (19), theregister.com (13), and patheos.com (13). 

As seen in Figure~\ref{fig:gpt-error}, we find that while at the beginning of 2022, there were seemingly no such error messages within our set of articles, by the end of April 2023, there were nearly six of these articles each day. We note that this graph also mirrors the behavior of the percentage of machine-generated articles that our DeBERTa detector found amongst all of our websites. Together, these results confirm that there \textit{has} been a noted increase in the use of synthetic content generation by our set of news websites in 2022 and 2023.
\begin{table}

\centering

\fontsize{6.4pt}{8}

\selectfont

\setlength\tabcolsep{4pt}

\begin{tabular}{llr|llr}

\textbf{Jan.} & & \textbf{CrUX}&\textbf{April} & & \textbf{CrUX} \\

\textbf{2022}&\textbf{\% Syn.} & \textbf{Rank} & \textbf{2023} & \textbf{\% Syn.} & \textbf{Rank} \\ \toprule

opensecrets.org & 42.5\% & $<$100K& china.org.cn & 34.9\% & $<$1M\\

theodysseyonline.com & 26.2\%& $<$1M& globaltimes.cn& 26.3\% & $<$100K\\

logically.ai & 17.2\%& $<$10M& thelist.comm& 26.0\% & $<$1M\\

china.org.cn& 16.3\%&$<$1M &bjreview.com & 26.0\%&  $>$10M+\\

globaltimes.cn& 16.0\%&$<$100K  &thefrisky.com & 23.6\% & $<$1M\\

egypttoday.com & 15.0\%&$<$1M  &northkoreatimes.com& 23.1\% & $>$10M+\\

sourcewatch.org & 14.4\%& $<$1M &egypttoday.com & 21.0\% &$<$1M\\

bleacherreport.com& 9.84\%& $<$100K& waynedupree.com& 20.1\%& $<$1M\\

thequint.com & 9.81\%& $<$10K & ancient-origins.net & 15.3\% & $<$100K \\

africanews.com& 9.64\%&$<$1M &entrepreneur.com & 15.0\% &  $<$100K\\

\bottomrule

\end{tabular}

\caption{\label{tab:largest-percentage-increase} Websites with the largest percentage of synthetic content (with at least 100 articles in that month) in January 2022 and in April 2023.}

\vspace{-10pt}
\end{table}

We finally note that we observe a small but noticeable dip in the percentage and amount of synthetic content in Figures~\ref{fig:gpt_over_time} (particularly among misinformation websites) and~\ref{fig:gpt-error} between February and March 2023. We find, as seen in Figure~\ref{fig:gpt_over_time}, that unreliable websites such as foreignpolicyi.org, prophecynewswatch.com, and awarenessact.com, in particular, drove the initial increase in machine-generated content in January and February 2022, before dramatically decreasing their amount of synthetic content in the following month. Examining Google trends data, we also observe that ChatGPT experienced a noticeable dip/decline (from 87\% of peak search traffic on February 5 to 75\% peak search traffic on March 5) in popularity in the United States during this period perhaps (but not definitively) explaining this small decline.


\vspace{3pt}
\noindent
\textbf{Trends Among Popular and Unpopular Websites.}
To understand how popularity and website size correlated with the near doubling of machine-generated content in 2022 and 2023, we plot the percentage of machine-generated/synthetic articles over time in Figure~\ref{fig:gpt_over_time} for websites within different rank buckets and striated by whether they are considered unreliable/misinformation or reliable/mainstream. As seen in Figure~\ref{fig:gpt_over_time}, there is a general upward trend in the amount of machine-generated articles across every popularity stratum.

Examining these increases within particular brackets of popularity, we see (as pictured in Figure~\ref{fig:gpt_over_time} and calculated in Table~\ref{tab:increase-brackets}) that the least popular websites saw the largest percentage increase in the use of synthetic articles. For both unreliable/misinformation and reliable/mainstream categories, we observe that for websites that rank $>$10M+ in popularity, the percentage of their articles that were synthetic increased by 3.42\% (736\% relative increase) and 2.13\% (423\%) on average, respectively. By contrast, among the most popular misinformation/unreliable websites (\textit{e.g.}, breitbart.com, zerohedge.com) and mainstream/reliable websites (\textit{e.g.}, cnn.com, foxnews.com), synthetic articles had a smaller 0.70\% (175\% relative increase) and a 0.36\% (27.0\%) increase overall. Indeed, calculating the websites with the most machine-generated content, we again observe in Table~\ref{tab:largest-percentage-increase} that the websites that had the largest amounts of synthetic content were all fairly small or unpopular small.


\vspace{3pt}
\noindent
\textbf{Topics Addressed by Synthetic Articles\label{sec:topics}.}
While misinformation websites and less popular websites have seen the largest increase in the use of synthetic articles, many reliable and large news websites also heavily use synthetic articles. However, as noted in Section~\ref{sec:related}, many reliable news sites have acknowledged their use of these machine-generated articles and utilize them in a benign manner. To understand different websites' use of synthetic articles, in this section, we analyze the topics addressed by synthetic articles among different types of websites and how this has changed between January 2022 and May 2023.

\begin{figure*}
\centering
\begin{subfigure}{0.48\textwidth}
  \centering
  \includegraphics[width=1\textwidth]{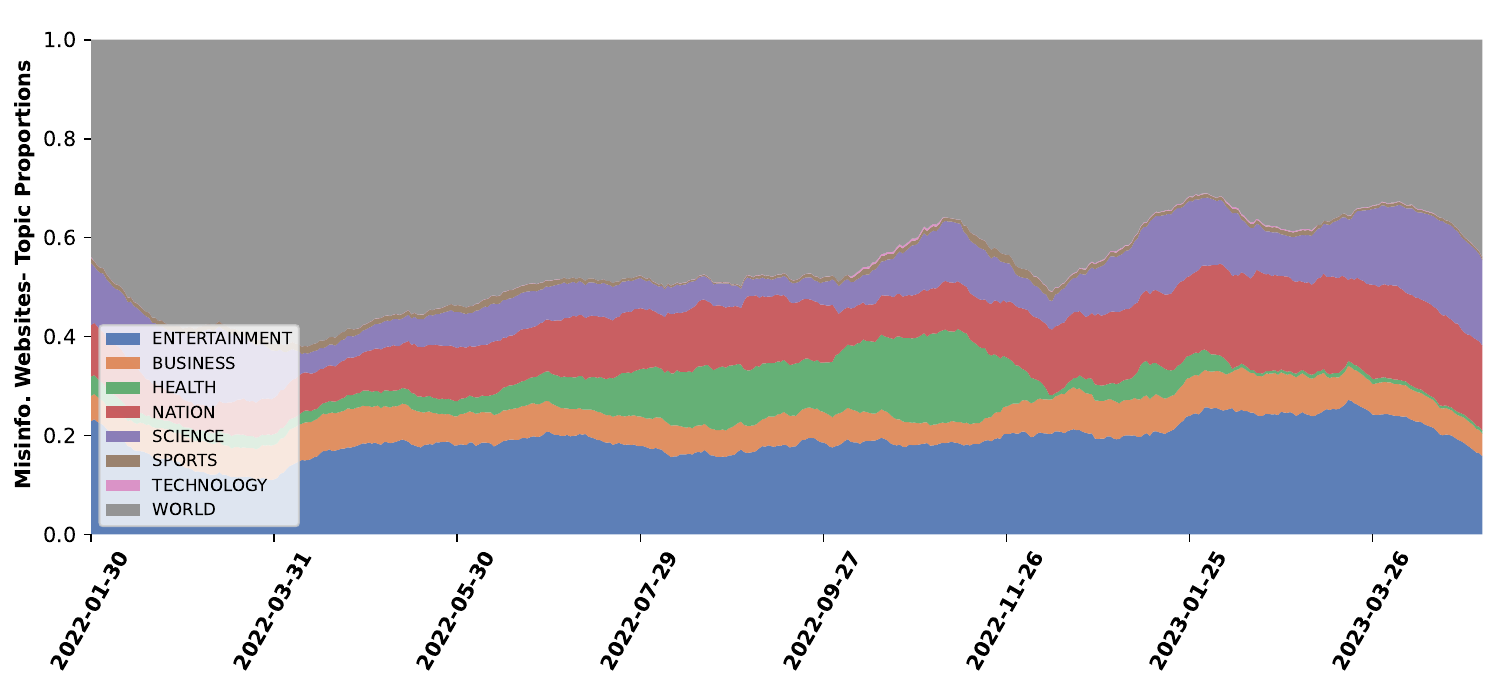}
  \caption{\label{fig:misinfo-synthetic-topics}Misinfo Synthetic Topics}

\end{subfigure}%
\begin{subfigure}{0.48\textwidth}
  \centering
  \includegraphics[width=1\textwidth]{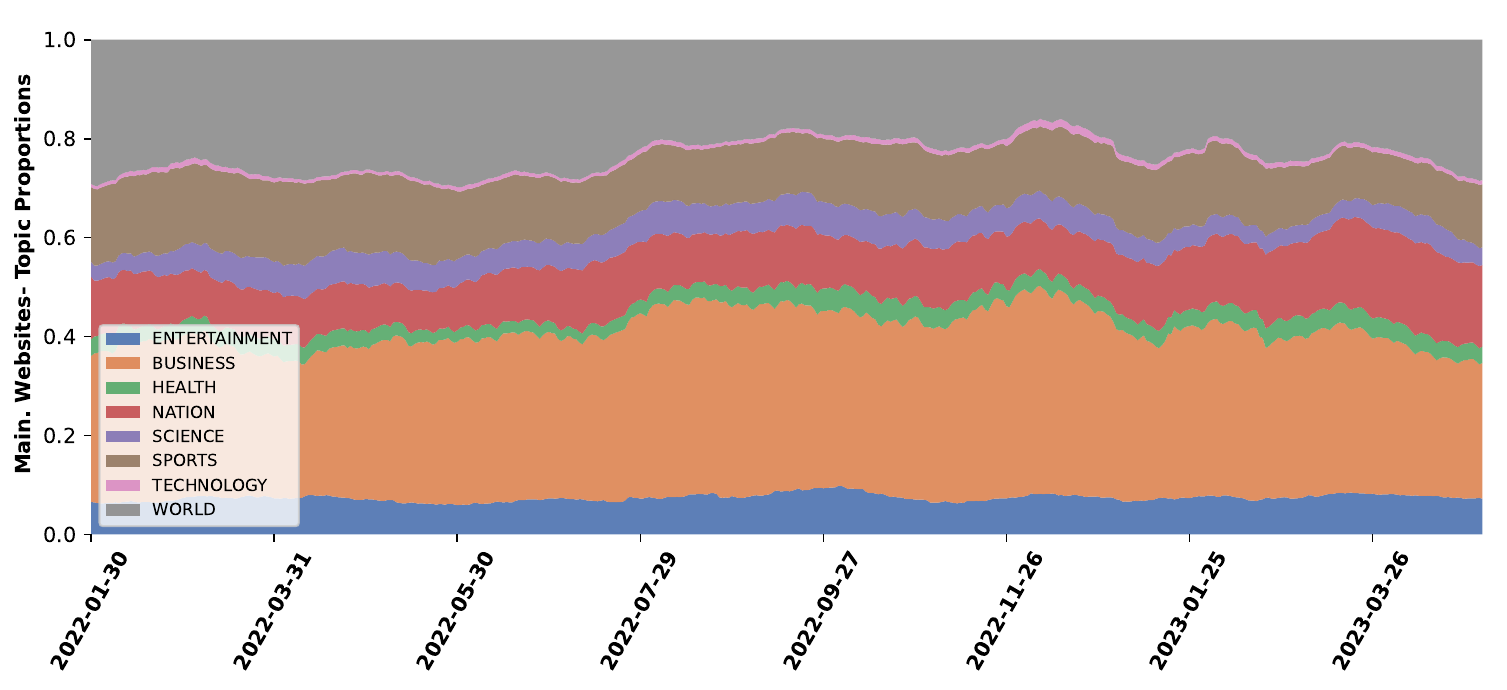}
    \caption{\label{fig:mainstream-synthetic-topics} Mainstream Synthetic Topics}
\end{subfigure}
\caption{The plurality of synthetic articles from mainstream/reliable websites is related to the \textit{Business } topic. In contrast, the majority of synthetic articles from misinformation/unreliable websites are related to \textit{Entertainment}.\textit{World} affairs, and \textit{US/Nation} current events.}
  \label{figure:conspiracies-to-different-types}
  \vspace{-10pt}
\end{figure*}


To identify the topics within our identified set of machine-generated articles, we train a DeBERTa-based classifier to identify the topic of an article based on its text. As training data, we utilize the News Catcher Topic Labelled dataset,\footnote{https://www.kaggle.com/datasets/kotartemiy/topic-labeled-news-dataset} which contains topic labels for 106,395 different articles as belonging to 8 different categories \textit{\{Business,
Entertainment,
Health,
US/Nation,
Science, 
Sports, 
Technology, and World\}}. We note while the original dataset only contained the title of each article, the dataset also included the original URL. As such, using the method outlined in section~\ref{sec:dataset}, we gather the set of articles listed in the dataset and subsequently train a DeBERTa-based classifier to correctly label articles based on their content. We note that a significant portion of these URLs were not available; as a result, we trained our model on a subset of 79,000 articles from the original dataset, further removing articles that were less than 1000 characters. Keeping out a 10\% of this dataset as a test dataset, upon training, we achieve a 0.819 $F_1$ score an average of 0.819 precision across the eight categories. Once trained, we finally categorize the topic of each of the 15.46 million articles within our dataset. 

\begin{table}

\centering

\fontsize{8.5pt}{8}

\selectfont

\setlength\tabcolsep{8pt}

\begin{tabular}{lr|lr}
Topic & Odds Ratio  & Topic &Odds Ratio\\\toprule
Entertainment & 0.68 &Science & 1.58  \\ 
Business & 0.61 & Sports &  0.23 \\ 
Health & 2.06 & Technology & 0.21 \\ 
Nation & 0.77 & World &  1.56\\ 
\end{tabular}
\caption{\label{tab:odds-misinfo}Odds Ratio for the amounts of \textbf{synthetic articles and human-written articles from misinformation websites} for each topic category.}
\vspace{-5pt}

\end{table}
Plotting the proportion of each topic amongst synthetic articles from misinformation websites, as seen in Figure~\ref{fig:misinfo-synthetic-topics}, a significant portion of synthetic articles from misinformation websites concerned \textit{World} affairs, \textit{Nation/US}-current events, \textit{Science}, and \textit{Entertainment}. For example, among our set of synthetic articles from misinformation websites, we identify a variety of articles about concerns about tensions between Russia and Ukraine, COVID-19 vaccines, and updates about the love life of Ed Sheeran. Calculating the odds ratio between the number of synthetic and human-written articles for each of our topic categories, as seen in Table~\ref{tab:odds-misinfo}, among our selection of misinformation websites, relative to their own topic proportions, misinformation websites were most likely to utilize synthetic articles for \textit{Health} and \textit{Science} related topics. This suggests that misinformation websites \textit{have} proportionally utilized synthetic articles for both mundane topics like Entertainment \emph{and} more serious topics such as Health~\cite{Peiser2019}.
\begin{table}

\centering

\fontsize{8.5pt}{8}

\selectfont

\setlength\tabcolsep{8pt}
\begin{tabular}{lr|lr}
Topic & Odds Ratio  & Topic &Odds Ratio\\\toprule
Entertainment & 0.59 &Science & 1.48  \\ 
Business & 1.53 & Sports &  0.85 \\ 
Health & 0.89 & Technology & 0.66 \\ 
Nation & 1.14 & World &  0.80 \\ 
\end{tabular}
\caption{\label{tab:odds-mainstream} Odds Ratio for the amounts of \textbf{synthetic articles and human-written articles from  mainstream websites} for each topic category. }
\vspace{-5pt}

\end{table}

Plotting the proportion of each topic amongst synthetic articles from mainstream websites, as seen in Figure~\ref{fig:mainstream-synthetic-topics}, the plurality of synthetic articles concern \textit{Business}. Indeed, as discussed previously, websites ranging from Bloomberg to Reuters have utilized synthetic articles to give updates on financial markets (Figure~\ref{figure:machine-generated}). Furthermore, again calculating the odds ratio between the number of synthetic and human-written articles for each of our topic categories, as seen in Table~\ref{tab:odds-mainstream}, among our set of mainstream websites, relative to their own topic proportions, mainstream websites are most likely to utilize synthetic articles for  \textit{Science} and \textit{Business} topics. This again reinforces prior reporting about the use of synthetic articles among mainstream websites. 

Finally, calculating the odds ratio (Table~\ref{tab:odds-misinfo-mainstream}) between the rates of usage of synthetic articles per category between mainstream and misinformation websites, we further observe that misinformation news and mainstream websites, throughout our period of study were more likely to utilize synthetic articles on topics related to \textit{Entertainment}, \textit{Health}, and \textit{Science}, and \textit{World} affairs. In contrast, mainstream websites were more likely to utilize synthetic articles for \textit{Business}, \textit{Technology} (very small proportion), and \textit{Sports}.  We observe similar proportions of \textit{US/Nation} topics between misinformation and mainstream websites. 

\begin{table}

\centering

\fontsize{8.5pt}{8}

\selectfont

\setlength\tabcolsep{8pt}



\begin{tabular}{lr|lr}
Topic & Odds Ratio  & Topic &Odds Ratio\\\toprule
Entertainment & 2.96 &Science & 1.91  \\ 
Business & 0.13 & Sports &  0.06 \\ 
Health & 1.67 & Technology & 0.11 \\ 
Nation & 1.02 & World &  2.75 \\ 
\end{tabular}
\caption{\label{tab:odds-misinfo-mainstream} Odds Ratio for the amounts of synthetic articles \textbf{between misinformation and mainstream websites} for each topic category. As seen above, misinformation websites are more likely to have synthetic articles about \textit{Entertainment}, \textit{Health}, \textit{Science}, and \textit{World}-related topics compared to mainstream websites. We observe similar proportions of US/Nation topics between misinformation and mainstream websites. }
\vspace{-10pt}

\end{table}

\vspace{3pt}
\noindent
\textbf{Estimating the Impact of ChatGPT.} As seen in the previous sections, misinformation websites and less popular websites saw the largest increases in the use of synthetic articles. In order to estimate how the introduction of ChatGPT specifically may have affected the levels of synthetic content on news websites, we now utilize an ARIMA model~\cite{zhang2003time} to perform an \textit{interrrupted-time-series} analysis. Namely, we examine whether there was a direct jump in the number of synthetic articles above expectation following the release of ChatGPT on November 30, 2022~\cite{OpenAI2022}.

As seen in Table~\ref{tab:arima-increase}, after the release of ChatGPT on November 30, 2022, we observe a noted jump (0.50\%) above expectation in the number of synthetic articles from misinformation websites. Many of the popularity ranking brackets of misinformation websites saw a statistically significant increase in the absolute percentage of their articles that were synthetic, with misinformation websites in the Rank $>$10M+ popularity bracket seeing the highest jump of 1.68\%. This was visually seen in Figure~\ref{fig:gpt_over_time}. We similarly observe that websites in every popularity bracket except those with Rank $>$10M+ saw the rate at which the percentage of synthetic articles increases, also increase (\textit{i.e.}, increase in the rate of increase). 

We further find that the groups of mainstream websites with popularity ranks $>$1M saw a marked increase in synthetic articles immediately following the release of ChatGPT on November 30, 2022. We further observe a trend increase for several mainstream website popularity brackets. Combined with our misinformation website results, this suggests that smaller, less popular, and otherwise less monitored websites were the ones that saw the biggest increase in synthetic articles following the release of ChatGPT. Indeed, the number of synthetic articles among \emph{all} groups of websites has been increasing and was at its highest levels on May 1, 2023 (Figure~\ref{fig:gpt_over_time}). We see this mirrored in the overall increase in the trend of mainstream websites' use of synthetic articles (increase in the rate of increase) in Table~\ref{tab:arima-increase}. We note that while this analysis is \emph{not} causal, it illustrates the noticeable increase in the percentage of synthetic articles among misinformation websites immediately following the release of ChatGPT.

\begin{table}

\centering

\fontsize{6.5pt}{8}

\selectfont

\setlength\tabcolsep{4pt}


\begin{tabular}{l|ll|ll}

& Misin. Abs.& Trend  & Main. Abs. & Trend\\

Rank &  \% Inc. & Inc. & \% Inc. & Inc. \\ \toprule

\textbf{All} &+$0.50\%^{***}$&+$0.006\%^{**}$ & +$0.04\%$ & +$0.001\%$\\ \midrule

Rank $<$ 10K & +$0.10\%^{***}$ &+$0.004\%^{***}$ & +$0.03\%$ & +$0.003\%^{***}$\\

10K $<$ Rank $<100$K & +$0.10\%$ &+$0.01\%$ & +$0.03\%$ & +$0.001\%$\\

100K $<$ Rank $<1$M & +$0.41\%^{***}$ &+$0.007\%^*$  & +$0.007\%$ & +$0.0005\%$\\

1M $<$ Rank $<10$M & +$0.12\%$ &+$0.006\%^{*}$ & +$0.19\%^{***}$ &+$0.002^{*}\%$ \\

Rank $>10$M & +$1.68\%^{***}$ & +$0.004\%$& +$0.79\%^{***}$ &+$0.004\%^{***}$ \\

\bottomrule

\multicolumn{5}{c}{ $^\ast p<0.05; \; ^{**} p<0.01; \; ^{***}p<0.001$} \\

\end{tabular}

\caption{\label{tab:arima-increase} Estimated absolute percentage increase immediately following the release of ChatGPT on November 30, 2022, in machine-generated articles (determined using an ARIMA-based interrupted time series analysis).}
\vspace{-15pt}

\end{table}

\section{Discussion and Conclusion}

In this work, we implement a DeBERTa-based model to classify 15.46~million articles from 3,074~news websites as \textit{human-written} or \textit{synthetic}. We find that between January 1, 2022, and May 1, 2023, the percentage of synthetic articles produced by mainstream/reliable news increased by 57.3\% while the percentage produced by misinformation/unreliable news websites increased by 474\%. Estimating the effect of ChatGPT, we observe a noticeable jump in the percentage of synthetic articles from misinformation websites and unpopular mainstream news around its release. We now discuss several limitations and implications of this work.


\vspace{3pt}
\noindent
\textbf{Limitations.}
We note that while we sampled our dataset from a large set of 3,074~news websites and gathered over 15.46M articles, we did not gather articles from \emph{every} news website and focused on English-language media. As such, our results largely do not apply to non-English media. Similarly, because we used pre-defined lists of misinformation websites, our work largely misses the \textit{probable} existence of new misinformation websites that appeared since the launch of ChatGPT\@.  



Because we take a conservative approach to our estimation of machine-\textit{generated/synthetic} texts and due to our removal of articles with characters lengths less than 1000 characters, the absolute numbers presented in this paper are only rough estimates of the percentage of articles on a given website that are machine-generated. As illustrated by Sadasivan et al.~\shortcite{sadasivan2023can}, reliable detection of these short texts is near impossible/largely impractical as large language models become more complex. As shown by Sadasivan et al.~\shortcite{sadasivan2023can}, as LLMs come to more closely match the distribution of written human language, the distinction between human-written and machine-generate texts disappears. As such, we note that while we manage to create a somewhat reliable detector in this work for longer articles for several released and public models, as more advanced and powerful models are developed, effective detection will be more difficult. Similarly, it has been shown that heavily human-edited machine-generated similarly are very difficult to detect as machine-generated~\cite{mitchell2023detectgpt} and in this work, we do not seek to detect these instances. As such, due to our conservative approach, our absolute percentage estimates are likely underestimates.

Furthermore, due to the limitations of our approach in building a model to \textit{estimate} the relative increase in machine-generated texts on news websites, our models are not universal classifiers for \textit{synthetic} texts. Most newspapers and outlets (as of early 2023), are not trying to purposefully evade AI detectors. Our models, which were trained on newspaper data from a given set of websites, are built for a particular context and cannot serve to universally detect synthetic texts.






\vspace{3pt}
\noindent
\textbf{Detection of Machine-Generated Media.}
We find that by training on data from a wide variety of generative models, we were able to outperform Open AI's released RoBERTa detector as well as several other released detectors~\cite{pu2022deepfake}. Furthermore, we find, as in prior works~\cite{gagiano2021robustness,pu2022deepfake}, that including data from common attacks can increase overall detection accuracy. We argue that future detectors applied to real-world data should account for these techniques.  

\vspace{3pt}
\noindent
\textbf{Small Websites and Synthetic Articles.} As seen throughout this work, while larger more popular websites have been slower to adopt the use of AI-generated and \textit{synthetic} content, smaller less popular websites in particular have shown the greatest relative increase in the use of synthetic (736\% increase among the least popular misinformation websites and 423\% increase among the least popular mainstream websites). We thus find that to fully understand the influence of synthetic media, as similarly argued by News Guard~\cite{newsguard2023}, researchers must document and study these less popular websites rather than just concentrating on the top and most frequently visited domains.

\vspace{3pt}
\noindent
\textbf{The Rise of Synthetic Misinformation.} We found that throughout 2022 and 2023, as large language models became more widely accessible, the percentage of machine-generated content on misinformation sites has had a 474\% relative increase. While at the beginning of 2022, a lower percentage of misinformation/unreliable news websites' content was synthetic (0.39\% vs. 0.88\%), we find that by May 2023, across all popularity brackets examined, misinformation websites had closed this gap (2.22\% vs. 1.39\%). Unlike popular mainstream websites, misinformation websites and unpopular mainstream websites experienced a noticeable jump in synthetic content after the release of ChatGPT (as determined by our \textit{interrupted-time-series} analysis). Furthermore, as shown by our topic analysis, misinformation websites have utilized these synthetic articles to address world affairs and health-related news more often than mainstream websites. While not every article posted on an unreliable/misinformation news website is necessarily misinformation, the rapid adoption of synthetic methods by misinformation websites for articles addressing world affairs and health news by these websites could have downstream negative effects. As such given the rapid adoption of the use of synthetic articles by misinformation and unpopular websites, in particular, we argue for future studies of how misinformation websites have utilized these technologies and how the content of these types of articles spread to social media and the broader Internet.


\bibliography{paper.bib}
\appendix
\section{Reddit Users and Synthetic Articles}

Having examined the increase in synthetic articles across the news ecosystem, we estimate whether Internet users are \emph{actually} interacting with this machine-generated content more. To do so, we analyze social media users from Reddit's overall interaction with synthetic articles.
\begin{table*}
\centering
\footnotesize
\begin{tabular}{lr|lr|lr}
Subreddit& \# Art. & Misinformation Domain &\# Art. & Mainstream Domain & \# Art. \\ \toprule
nofilternews & 984 & dailymail.co.uk & 96 & investopedia.com & 500\\
autotldr & 948 & globaltimes.cn  & 34 & reuters.com & 392\\
newswall & 442 & justthenews.com & 10 & cnbc.com & 391\\
stocks  & 201 & washingtonexaminer.com & 9 & newsweek.com & 161 \\
backfieldvacio & 142 & express.co.uk &7& bleacherreport.com& 153 \\
\bottomrule

\end{tabular}
\caption{\label{tab:subreddit-most-syn} Subreddits, misinformation websites, and mainstream websites with the most \textit{synthetic} Reddit submissions. As seen above, the subreddits and news sites with the most posted articles largely concern money/business, sports, and daily updated news reports. } 
\vspace{-10pt}
\end{table*}

\vspace{3pt}
\noindent
\textbf{Reddit Dataset.} To understand Reddit users' interaction with synthetic content, we first gather \emph{all} Reddit submissions/posts and their associated metadata (\textit{e.g.}, date posted, the subreddit of the post, number of comments, \textit{etc...}) that referenced an article published between January 1, 2022, and April 1, 2023, from one of the news websites in our dataset. To collect this Reddit data, we rely upon Pushshift~\cite{baumgartner2020pushshift}, which keeps a queryable replica of Reddit data, getting all available submissions posted between January 1, 2022, and March 31, 2023. Altogether, we identify a set of Reddit submissions that make use of 254,302 news articles (19,299 from misinformation/unreliable websites; 235,003 from mainstream/reliable news websites) from our list of URLs from 2022 and 2023. Among these articles, we identified 3,428 synthetic articles from mainstream websites and 209 synthetic articles from misinformation websites. We report the subreddits and news websites with the most synthetic articles that appear in our Reddit dataset in Table~\ref{tab:subreddit-most-syn}. We note that despite these low numbers, we perform this analysis to get an understanding of the broad trends in interaction with synthetic media on Reddit.

\noindent
\paragraph{Ethical Considerations.}
We collect only publicly available data from Reddit. In addition, we did not attempt to deanonymize any Reddit user. We note that we utilize public Pushshift data that was collected before Reddit ultimately blocked Pushshift on May 1, 2023, for breaking their terms of service. 
\begin{figure}
\centering
 \begin{subfigure}{.4\textwidth}
 \centering

  \includegraphics[width=1\linewidth]{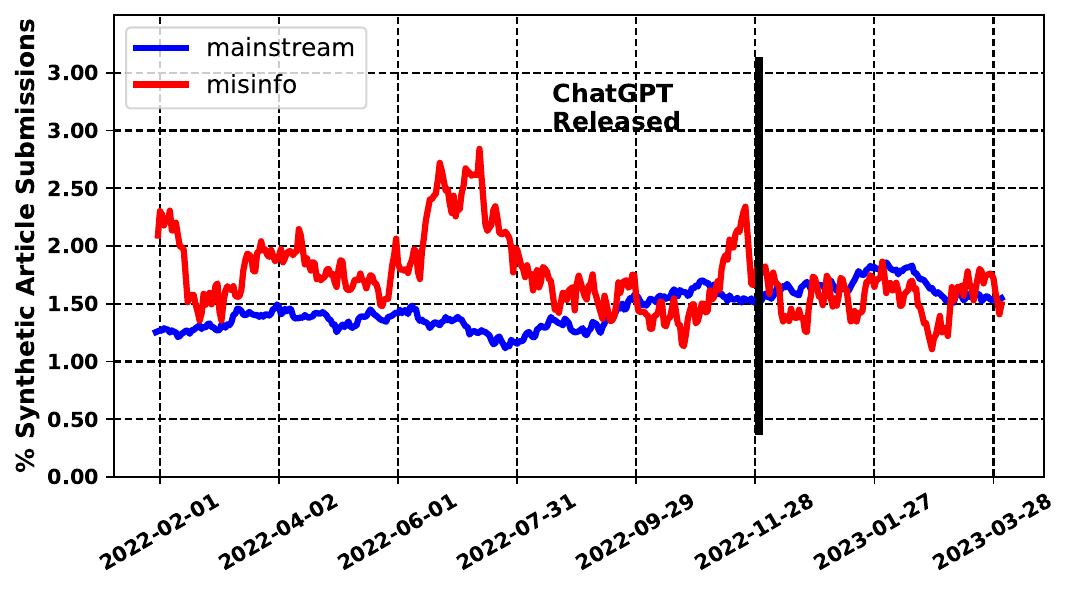}
 \end{subfigure}%
\caption{Percentage of Reddit submissions featuring news articles published between January 1, 2022, and March 31, 2023, that went to synthetic/machine-generated articles.\label{fig:reddit-submissions}}
\vspace{-10pt}
\end{figure}

\begin{figure}

 \centering

\begin{subfigure}{.40\textwidth}

\centering

 \includegraphics[width=1\linewidth]{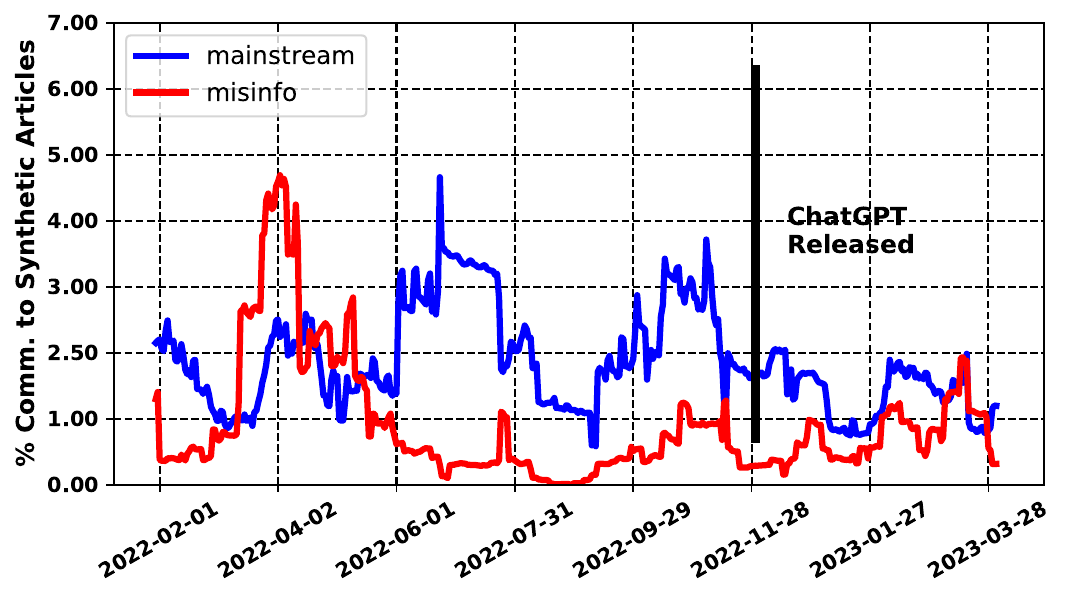}

\end{subfigure}%

\caption{Percentage of Reddit comments on news submissions between January 1, 2022, and March 31, 2023, that went to synthetic/machine-generated articles submissions. \label{fig:redditcomments} }

\vspace{-10pt}

\end{figure}

\vspace{3pt}
\noindent
\textbf{User Interaction with Synthetic Articles.} In order to examine how Reddit users have interacted with synthetic articles, we plot the percentage of synthetic articles among Reddit submissions that featured an article from our 3,074 news websites (Figure~\ref{fig:reddit-submissions}). In addition, among the submissions that hyperlinked to an article from our news article dataset, we determined the percentage of Reddit comments that were on Reddit submissions that featured a synthetic article rather than a human-written one (Figure~\ref{fig:redditcomments}).

\textit{Mainstream Synthetic Articles.} As seen in Figure~\ref{fig:reddit-submissions}, despite the overall relative percentage increase (57.3\%) in the use of synthetic new articles by mainstream/reliable news websites, we do not see a correspondingly large percentage increase in the percentage of mainstream/reliable news submissions on Reddit that are synthetic. Between January 1, 2022, and March 31, 2023, we only observed a slight increase from 1.26\% to 1.43\% (a 12.9\% relative increase). However, while synthetic articles in terms of proportions did not increase significantly among posted mainstream articles, in terms of raw numbers, we find that the daily average of mainstream synthetic article submissions went from 4.1 in January 2022 to 7.3 in March 2023 (78.0\% increase). Similarly, we observe a raw increase in the average daily number of comments on these submissions from 329.9 comments to 3,134.0 comments, an 850\% increase. However, as seen in Figure~\ref{fig:redditcomments}, in terms of the percentage of comments that went to submissions that featured synthetic articles, this corresponded to an absolute percentage decrease of 1.57\% (this amount has varied dramatically however).

We further determine whether these machine-generated articles tend to receive more or less interaction from Reddit users. Performing a pairwise comparison on a domain basis of the number of comments on human-written articles against synthetic articles, we determine, we find, after controlling for the particular website, that \textit{on average} synthetic articles from mainstream websites tend to receive approximately 5.83 fewer comments from Reddit users than human-written articles.\footnote{We apply a Mann–Whitney U-test and find this difference to be statistically significant (\textit{i.e.}, $p\approx0$).} Thus, while there has been a slight increase in the percentage of synthetic articles on Reddit from our set of mainstream websites, machine-generated articles from these mainstream/reliable websites were \textit{on average} less popular on Reddit than human-written articles.

\textit{Misinformation Synthetic Articles.} As seen in Figure~\ref{fig:reddit-submissions},  the percentage of Reddit submissions that feature synthetic news articles from misinformation/unreliable news websites remained relatively stable. We observe only an increase from 2.50\% in February 2022 (there was a temporary uptick in January 2022), to 1.67\% in March 2023, a 6.39\% relative increase. However, in terms of raw numbers, we observe that while on average 0.4 synthetic articles were featured in Reddit submissions each day in February 2022, this increased to an average of 0.5 in March 2023, a 20\% increase. This corresponds to a similar increase in the number of comments on these submissions. Specifically, we observe a raw increase in the average total number of comments on these submissions each day from 9.60 comments to 13.67 comments, a 42.6\% increase (Figure~\ref{fig:redditcomments}).

We again determine the difference in the amount of users' comments on synthetic and human-written articles. After controlling for the particular website, human-written content from our set of misinformation/unreliable news websites tends to receive approximately 5.61 more comments than synthetic content.\footnote{We again apply a Mann–Whitney U-test and find this difference to be statistically significant (\textit{i.e.}, $p\approx0$).} This shows, again, on the whole, that human-written articles \emph{tend} to see more engagement.

From this analysis, examining both mainstream and misinformation websites, we thus see that while there has been an uptick in the percentage of Reddit submissions that feature synthetic articles, there has not been a corresponding proportional increase in the percentage of Reddit comments interacting with these submissions.

\end{document}